\documentclass[lettersize,journal]{IEEEtran}
\usepackage{amsmath,amsfonts}
\DeclareMathOperator*{\argmin}{arg\,min}
\usepackage{algorithmic}
\usepackage{algorithm}
\usepackage{array}
\usepackage[caption=false,font=normalsize,labelfont=sf,textfont=sf]{subfig}
\usepackage{textcomp}
\usepackage{stfloats}
\usepackage{url}
\usepackage{verbatim}
\usepackage{graphicx}
\usepackage{cite}
\usepackage{xcolor}
\usepackage{enumitem}
\usepackage{amssymb}
\usepackage{multirow}
\begin{document}
\title{Enhancing Urban VANETs Stability: A Single-Hop Clustering Strategy in Metropolitan Environments}

\author{Pouya Firouzmakan, and Suprakash Datta
\thanks{This paper was produced by the IEEE Publication Technology Group. They are in Piscataway, NJ.}
}

\markboth{Journal of \LaTeX\ Class Files,~Vol.~14, No.~8, February~2024}{Journal of \LaTeX\ Class Files,~Vol.~14, No.~8, February~2024}%

\maketitle

\begin{abstract}
Vehicular Ad-hoc Networks (VANETs), a subclass of Mobile Ad-hoc Networks (MANETs), are expected to play a crucial role in the future of intelligent transportation systems (ITSs). A key objective of VANETs is to enable efficient and cost-effective communication among vehicles while supporting a large number of network participants and minimizing infrastructure dependency. However, the highly dynamic nature of vehicular networks poses significant challenges to their deployment. Clustering techniques are employed to address these challenges, with a strong emphasis on stability, as they directly influence the routing process and enhance the quality of service (QoS). This paper explores the feasibility of reducing reliance on roadside units (RSUs) in metropolitan areas while improving cluster stability. We propose an efficient clustering algorithm tailored for urban environments, leveraging existing metropolitan infrastructure to compensate for the absence of RSUs. Our approach designates public transportation buses as primary cluster heads (CHs), minimizing reliance on additional infrastructure, while stand-alone vehicles (SAVs) dynamically select additional CHs. Through comprehensive case studies and comparative analysis with existing algorithms, our results demonstrate the superior performance of the proposed method across different transmission ranges (TRs).
\end{abstract}


\section{Introduction}

\IEEEPARstart{V}ehicular Ad Hoc Networks (VANETs) have evolved from Mobile Ad Hoc Networks (MANETs) principles to cater specifically to vehicles. The primary concept is to leverage the increased number of vehicles and their potential for communication for emergency purposes, convenience, and message transmission~\cite{Kokare-Kakkar2021, Sarkar-Chakrabarty2021, Zhang-Ge, Bi-Shan, Ren-Zhang}. These applications are particularly beneficial in metropolitan environments. However, implementing VANETs in practice presents significant challenges, including frequent topology changes, reliable routing, stability, and security~ \cite{Abbas-Abdulsattar2022, Sarkar-Chakrabarty2021, Saleem-Zhou2019, Ferng-Abdullah2019}. 
These issues stem from the high mobility of nodes within VANETs, which also impacts the addressing mechanism in the routing process, leading to a substantial increase in update frequency. Additionally, similar to MANETs, some nodes may be unwilling to participate in the network. To address these challenges, Roadside Units (RSUs) can be deployed in urban areas to enhance VANETs~\cite{Liang-Wang, Jain-Jeyakumar2016, Yeung-Hui, Jalooli-Song, Benkirane}. The deployment of RSUs typically relies on multi-objective optimization frameworks aimed at minimizing costs, data transmission delays, and losses while maximizing area coverage and Quality of Service (QoS)~\cite{Liang-Wang, Zhang-Li2021, Kim-Velasco}.

However, transforming the network into connected clusters based on various criteria and approaches can be the most effective and efficient way to significantly reduce the complexity of the aforementioned challenges~\cite{Jabbar-Trabelsi2022}.
Clustering helps manage the impacts of topology changes, such as stability and reliable routing, within a much simpler framework. Additionally, it results in fewer IP address updates and simplifies routing algorithms. Notably, clustering in urban areas offers substantial advantages to VANETs, as congestion and traffic exhibit more intermittent behaviors.
 
Existing research highlights the importance of clustering for VANETs, especially in metropolitan areas. This paper introduces a single-hop metropolitan zone-based clustering algorithm (SMZCA). This algorithm enables vehicles to select cluster heads (CHs) and form new clusters, reducing the frequency of cluster changes. We have used an area zoning method to establish general vehicle directions, significantly enhancing cluster stability. Additionally, public transportation buses are designated as default CHs due to their exceptional ability to function as dynamic road units (DRUs) in routing protocols, which will be discussed in our future studies. The contributions of this paper are as follows:

\begin{itemize}
\item
We present a new zoning area approach to determine vehicles' general moving directions for integration into CH selection. Our method reduces complexities related to movement similarity factors and organizes the area to facilitate a clustering algorithm in large cities. To the best of our knowledge, no existing publications have focused on a comparable procedure.
\item
In addition to the algorithm for CH selection, we introduce a decentralized algorithm to create clusters for stand-alone vehicles (SAVs). These proposed algorithms, based on area zoning, have demonstrated significant improvements in cluster stability without relying on RSUs in large cities.
\item
Rather than using a routing protocol that may not adapt well to any clustering approach, especially for comparison purposes, we have proposed a new centralized stability factor. This factor evaluates clustering algorithms based on the number of clusters that vehicles join or leave within the area. 
\item
To showcase the capabilities of the proposed clustering algorithm, we investigated two case studies to provide insightful comparisons regarding the presence of RSUs. Additionally, we conducted a comprehensive comparison with the clustering algorithm proposed in \cite{arkian-Atani}, which is based on a factor called BeFit, and the dynamic single-hop clustering algorithm (DSCA) \cite{Katiyar-Gupta}, developed on top of the BeFit factor. The results demonstrate the superior ability of the proposed clustering algorithm to enhance cluster stability. 
\end{itemize}
The remainder of this paper is organized as follows. Section II summarizes related works. Section III presents the framework to prepare the area for the proposed clustering algorithm implementation. Section IV focuses on the proposed clustering algorithm. Section V introduces evaluation criteria to assess cluster stability. Section VI discusses the proposed clustering algorithm's capabilities through comparisons based on the introduced evaluation criteria and obtained results. Finally, Section VII draws the conclusion.

\section{Related Work}
A clustering framework can streamline the computation of metropolitan environment properties, resulting in more reliable, stable, and scalable routing~\cite{Khan-Abolhasan2018}. Additionally, it significantly reduces dependency on infrastructure, such as RSUs. However, the design of a clustering algorithm is crucial to meet the minimum requirements and demands. As our focus is on single-hop clustering in metropolitan areas in this study, relevant works have been reviewed in this section.

\subsection{Clustering in VANET}
In recent years, researchers have focused on developing clustering solutions for VANETs. These solutions include meta-heuristic algorithms, distribution algorithms, fuzzy logic, weighted algorithms, and other modified algorithms \cite{Kokare-Kakkar2021}. Several studies have proposed clustering methods aimed at enhancing stability and performance in urban VANETs \cite{Abbas-Abdulsattar2022, Saleem-Zhou2019, Ferng-Abdullah2019, Affandi-Suardi2021, Cheng-Yuan2020, Alsuhli-Khattab2019, Shah-Habib2018}. To reduce the number of clusters in urban areas, \cite{Abbas-Abdulsattar2022} introduced a hybrid CH election method. A fuzzy CH selection method is considered in \cite{Saleem-Zhou2019}. \cite{Ferng-Abdullah2019} proposed and evaluated a clustering scheme for urban VANETs simulated in NS-3, where clusters are formed based on metrics such as signal-to-noise ratios (SNRs), packet-reception ratios (PRRs), and speeds to estimate link quality. Vehicles send "HELLO" packets to collect data and receive relevant information from other vehicles to calculate these metrics. In \cite{Affandi-Suardi2021}, the authors used SUMO and OMNET++ simulators to address reliable communication among medical vehicles, considering various communication types like vehicle-to-vehicle (V2V), vehicle-to-infrastructure (V2I), and vehicle-to-pedestrian (V2P). This study employed a weighted clustering algorithm to account for mobility, distance, and connectivity. Additionally, several papers have utilized learning models for clustering \cite{Cheng-Yuan2020, Alsuhli-Khattab2019}. \cite{Cheng-Yuan2020} proposed dynamic clustering based on connectivity prediction using a recurrent neural network (RNN), highlighting user QoS. \cite{Alsuhli-Khattab2019} used an Online Sequential Extreme Learning Machine (OS-ELM) to continuously learn and predict vehicle behavior around intersections, ensuring clustering stability. This learning model is not deep, making the learning process efficient. To achieve stable and connected clustering, \cite{Shah-Habib2018} proposed an optimization algorithm based on the Moth-Flame Optimizer (MFO), considering the number of clusters. Promising results have also been seen in using public transportation as a feature, although not many studies have focused on it \cite{Tseng-Wu2020}. \cite{Tseng-Wu2020} proposed a clustering algorithm based on buses and their regulated routes to improve cluster stability.

\subsection{Single-hop communication paradigm}
In the context of single-hop clustering and related communication paradigms, the authors in \cite{Katiyar-Gupta} proposed a dynamic clustering scheme that focuses on CH selection and cluster maintenance. They introduced a suitability index for vehicles to select CHs based on BeFit and connectivity factors, which are calculated using road length, vehicle length, and speed. By categorizing the changes affecting cluster stability into internal and external factors—reflected by the change rate in cluster members (CMs) and the number of common nodes between clusters, respectively—\cite{Abboud-Zhuang} conducted a stochastic stability analysis. To address external stability, the authors determined the distributions of time before the first change in the cluster-overlap state and the interval between two related changes in neighboring clusters, applying a cluster-overlap state analysis. This analysis uses a discrete-time Markov chain, with the cluster-overlap state indicated by the distance between two CHs. A similar procedure was used for internal stability analysis, considering the time before the first change in cluster membership and the interval between two changes in CMs. It is important to note that this reference assumed the use of certain clustering algorithms to form the clusters.

\subsection{Zone-Based VANETs}

In VANETs, zones have primarily been used for routing, particularly geographical routing \cite{Wang-Mitton2023, Lin-Kang2017}, rather than clustering. However, there is a lack of studies that integrate mixed area zones with clustering to align better with geographical or other routing algorithms. In \cite{Wang-Mitton2023}, the authors addressed broadcast storms in VANETs by proposing a preferred-zone broadcasting protocol. This protocol detects neighboring zones and allows nodes within them to rebroadcast received data. Due to the limited availability of RSUs for routing protocols, \cite{Lin-Kang2017} proposed a routing protocol based on moving zones, which are formed based on the movement similarities of vehicles. Additionally, the authors assumed that vehicles are equipped with onboard units (OBUs) to handle networking and computational tasks.

\section{Area Preparation and Zone Search Strategy}\label{sec: area-preparation}
\begin{figure}[htp]
\centering
\includegraphics[width=3in, height=2.0in]{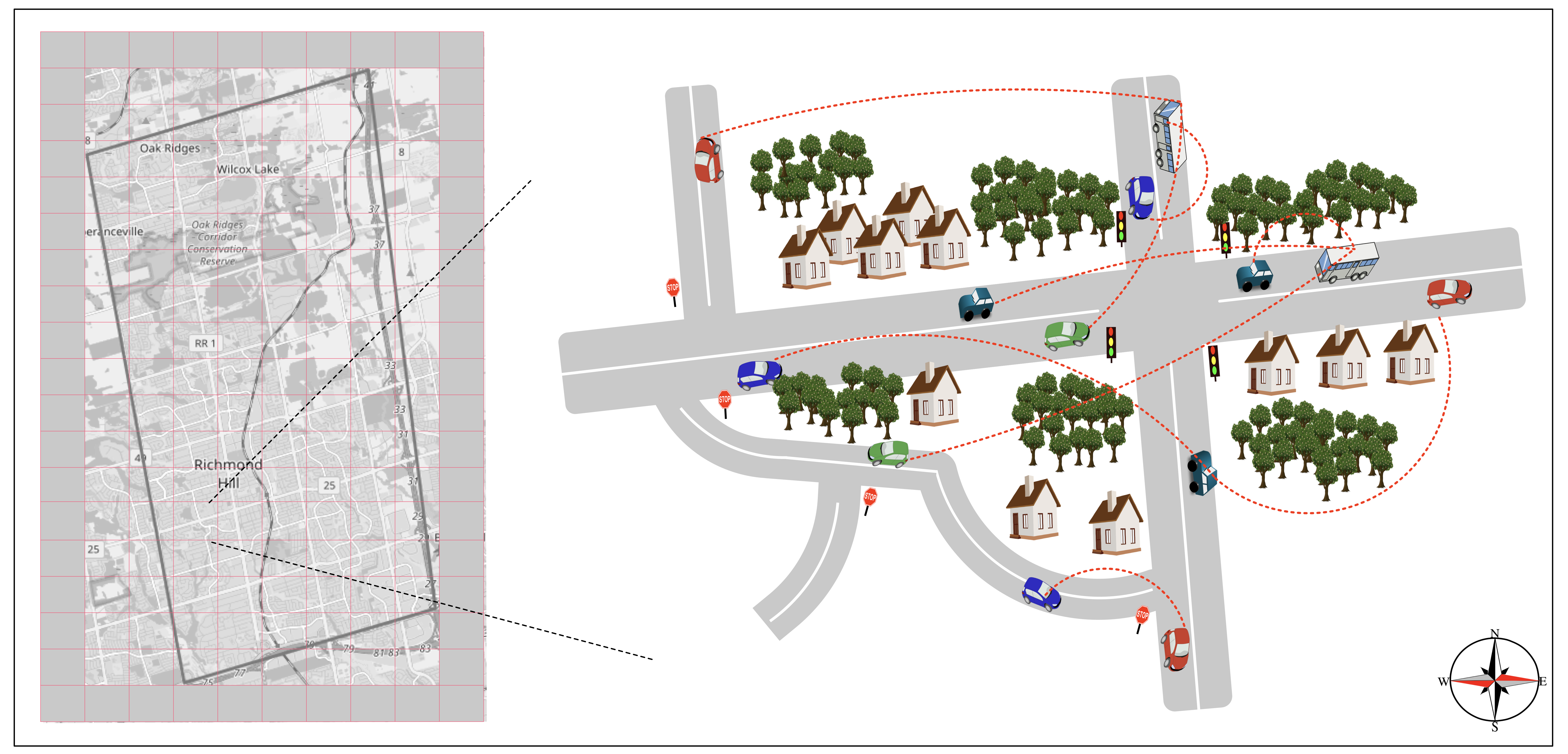} 
\caption{Area-zoning scheme for VANETs in metropolitan area}
\label{fig_1}
\end{figure}

This section explains the approach for translating the area to vehicles. Before addressing the clustering stage, dividing the area into smaller zones enhances the vehicles' understanding, enabling them to make efficient decisions. Figure \ref{fig_1}, which pertains to Richmond Hill City as part of the Greater Toronto Area (GTA), illustrates the general area preparation framework as an example. The proposed framework includes assigning zones to the area, as well as zone search and determination. In practice, many parts of the GTA are considered metropolitan areas individually. Therefore, this framework would be applicable across all metropolitan areas, especially those in close proximity to each other.

\subsection{Padded Zone Assignment to the Metropolitan Area}
Our proposed area zoning method enhances the clustering strategy in VANETs and streamlines the IP addressing process, simplifying future routing protocols. To achieve this, we consider a rectangle encompassing the minimum and maximum latitude and longitude of the metropolitan area, which is then divided into equal rectangular zones. These zones can be adjusted in size based on the vehicles' common TRs, either larger or smaller. Ideally, each zone would have equal length and width, but in other cases, they will be nearly equal. In this context, space complexity becomes a trade-off factor for short TRs. The number of zones, $n_z$, can be calculated as follows:
\begin{align} 
\label{num_zones} 
n_z = n_{r}*n_{c}
\end{align} 
where,
\begin{align} 
\label{num_cols}
n_{r} &= \lfloor \alpha^{-1}W_a \rfloor\\
\label{num_rows}
n_{c} &= \lfloor \alpha^{-1}L_a \rfloor
\end{align} 
In \ref{num_zones}-\ref{num_rows}, $n_r$ and $n_c$ represent the total number of zone columns and rows assigned to the area, respectively. $W_a$ and $L_a$ denote the length and width of the area in kilometers (km). The scaling factor $\alpha$ adjusts the size of the zones based on the common TR of vehicles. Consequently, the length ($l_z$) and width ($w_z$) of each zone are given by \ref{l_zone} and \ref{w_zone}, respectively.
\begin{align} 
\label{l_zone}
w_z &= {n_r^{-1}}W_a \\
\label{w_zone}
l_z &= {n_c^{-1}}L_a
\end{align}
As shown in Fig. \ref{fig_1}, the shaded zones are included as two rows and columns (to maintain the simplicity). These zones provide opportunities for incoming vehicles to update traffic information and protocols, initiating their participation in the metropolitan VANET. The data is primarily broadcasted to vehicles within the area by DRUs. Additionally, these zones benefit the routing process for vehicles carrying packets and intending to leave the area. Depending on each vehicle's direction, these zones also handle similar responsibilities for other areas if they are not part of other zoned areas. 

To define the zones, each one is determined by its minimum and maximum latitudes and longitudes, specifically its southwest and northeast points~\cite{Kitsis-Datta2018}. The zones are numbered starting from 0, beginning in the southwest and moving eastward, then northward.

\subsection{Zone-Search and Determination}
To leverage the benefits of zoning, vehicles must be able to identify their current zones quickly enough to prevent clustering challenges caused by rapid network changes. Since each vehicle and DRU needs to periodically determine its zone, an efficient search mechanism is essential. While zone determination can be directly performed using latitude and longitude in uniform zoning, this method does not apply to non-uniform zoning, where zone sizes vary. Although non-uniform zoning is advantageous for routing due to flexible region sizes, uniform zoning offers a more consistent representation of vehicle movement direction, making it more suitable for clustering and CH selection.

\begin{figure}[t]
\centering
\includegraphics[width=3.5in, height=2.25in]{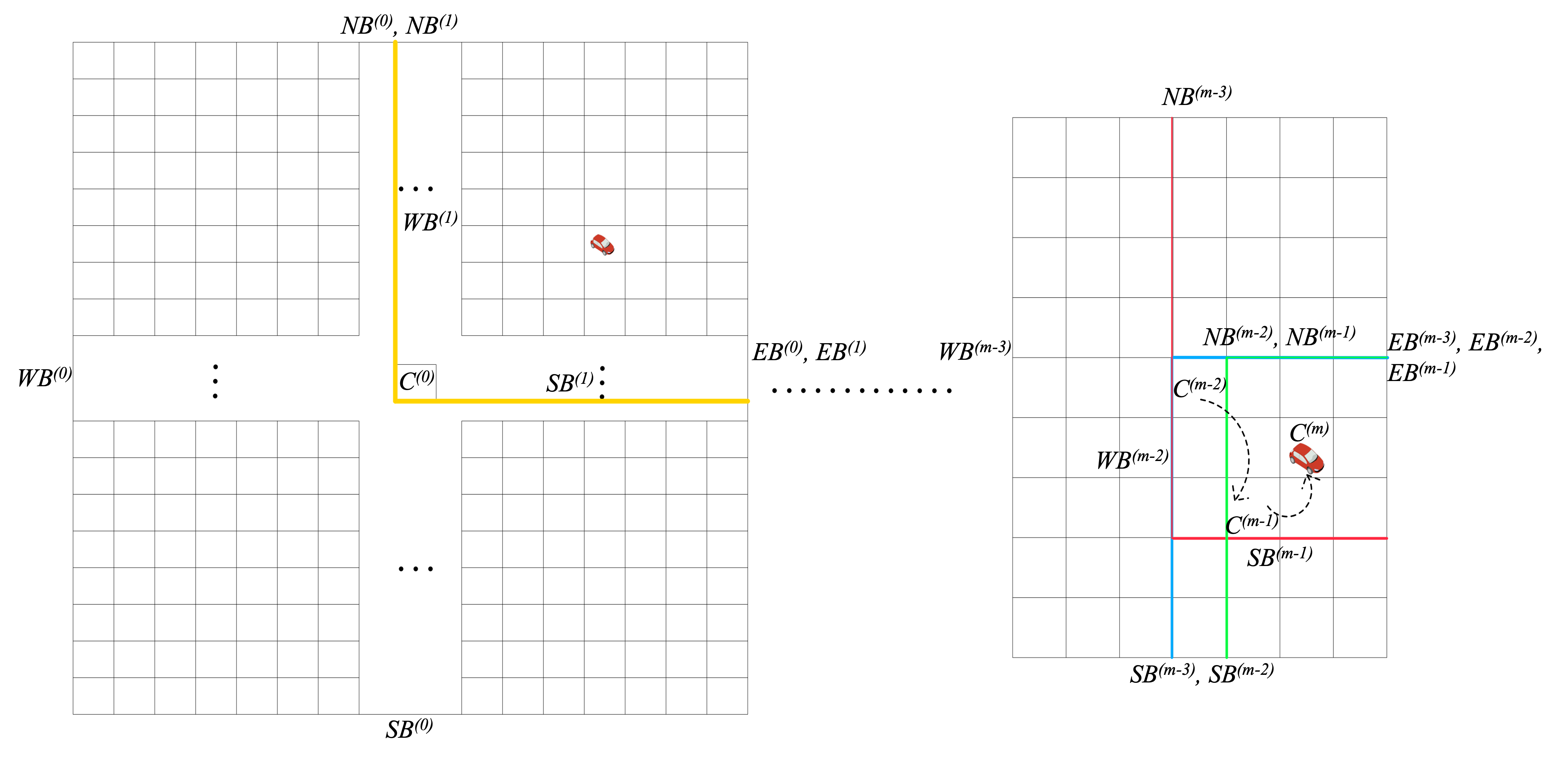} 
\caption{Zone search strategy used by vehicles}
\label{fig_2} 
\end{figure}

In this paper, we propose a divide-and-conquer approach for zone determination to ensure generality and consistency, achieving $O(\log{n_z})$ time complexity for each query. As shown in Fig. \ref{fig_2}, this method dynamically updates zone $C$, and iteratively refines the search by comparing the vehicle’s latitude and longitude, obtained from the global positioning system (GPS), with the zone’s boundaries until convergence. The vehicle updates the north boundary (NB), south boundary (SB), west boundary (WB), and east boundary (EB) to find a new and closer zone. This approach maintains a unified methodology that supports both uniform and non-uniform zoning, ensuring adaptability for future routing extensions while preserving efficiency. In the zoned area, these boundaries correspond to the lower and upper rows and columns. Therefore, $C$ is calculated and updated using \ref{centre_zone_id}.

\begin{align} 
\label{centre_zone_id} 
C^{(m)} &= n_{c}\left(\beta^{(m)}_{ns}-1\right) + \left(\beta^{(m)}_{ew}\right)-1.
\end{align} 
where,
\begin{align} 
\label{beta_ul} 
\beta^{(m)}_{ns} &= \left\lceil \frac{NB^{(m)} + SB^{(m)}}{2}\right\rceil\\
\label{beta_rl} 
\beta^{(m)}_{ew} &= \textit{round}\left(\frac{EB^{(m)}+WB^{(m)}}{2}\right)
\end{align}

Therefore, to enhance clustering efficiency, even the most basic technology package for vehicles should include GPS. It is important to note that, in addition to the current zone, each vehicle also saves its previous zone, unless it has remained in the same zone for an extended period or was turned on in the same zone. The zone search strategy is detailed in Algorithm \ref{zone_serach_algorithm}.

\begin{algorithm}[H]
\caption{Zone Search and Determination.} 
\label{zone_serach_algorithm}
\begin{algorithmic}[1]
\STATE \textit{input predefined zones as well as} $(v_{lat}, v_{long})$
\STATE $m \gets 0$ 
\STATE \textit{initiate} $NB^{(m)}, SB^{(m)}, WB^{(m)},$ \textit{and }$EB^{(m)}$
\STATE \textit{initiate} $C^{(m)}$
 
\WHILE{True}
    \IF{\textit{vehicle in the $C^{(m)}$'s zone}}
        \STATE \textit{vehicle updates its zone-ID}
        \STATE \textit{break}   
    \ELSE
        \STATE $m \gets m+1$ 
        \STATE \textit{update} $NB^{(m)}, SB^{(m)}, WB^{(m)},$ \textit{and }$EB^{(m)}$
        \STATE \textit{update} $C^{(m)}$
    \ENDIF  
\ENDWHILE
\end{algorithmic}
\end{algorithm}

\section{Proposed SMZCA}{\label{sec: proposed SMZCA}}
This section describes the proposed algorithms for creating or joining clusters for vehicles. Since the priority is to have fewer clusters, simpler routing, and minimize hops between the source/sender and destination/receiver, the CH selection approach is defined first. Next, the method for the remaining SAVs to create a new cluster and select CHs among themselves is specified. Finally, the maintenance procedure is briefly explained. 

\subsection{CH selection algorithm}

In this paper, the CH selection approach used by SAVs is based on weighted factors, collectively termed the CH eligibility criterion (CHEC). CHEC is defined by vehicle direction, speed, and distance, with a particular emphasis on direction in this study. Consequently, all CMs and SAVs continuously broadcast their information using a beacon-based mechanism. The features shared by node (vehicle) $i$ for clustering purposes are as follows:
\begin{equation}
\label{features}
\begin{aligned}
\textit{feat}_i = \{\textit{id}_i, \textit{loc}_i, \textit{z}_i, \textit{pz}_i, \overrightarrow{s}_i, \textit{TR}_i, \textit{ch}_i, \textit{cm}_i\}
\end{aligned}
\end{equation}
where the term '\textit{id}' refers to the unique identification ID assigned to the vehicles and DRUs by the owner or public transportation company during the registration process, '\textit{loc}' is the location, including its latitude and longitude, '\textit{z}' is the current zone determined using Algorithm \ref{zone_serach_algorithm}, '\textit{pz}' refers to the previous zone, '$\overrightarrow{s}$' is the speed, and '\textit{TR}' is the maximum TR. Additionally, the boolean parameters '\textit{ch}' and '\textit{cm}' (with a value of 1 for 'yes' and 0 for 'no') serve to indicate various aspects of the entity denoted by \textit{i}. Specifically, '\textit{ch}' indicates whether a node is a CH, and '\textit{cm}' signifies its membership status within a cluster. Additionally, the acknowledgements (ACKs) sent by CHs and DRUs include another feature, '$d$', a boolean parameter that indicates whether the CH is a DRU. Sending an ACK also signifies that the CH has capacity for new CMs, up to a maximum number. The maximum number of CMs in a cluster varies depending on whether the CH is a DRU. It should be noted that CHs, like CMs and SAVs, periodically send beacons to establish connections among themselves, thereby increasing scalability and expanding the network.

 Meanwhile, SAVs update their memory regarding nearby CHs, DRUs, CMs, and other SAVs that have sent back ACKs ($v_{bACK, i}$, $v_{chACK, i}$, $v_{cmACK, i}$, and $v_{saACK, i}$, respectively) using the shared boolean information. Additionally, they define a vector called the relative feature vector ($\overrightarrow{RFV}$) for each ACK received from a CH to select the most suitable option among the available potentials, as follows: 
\begin{equation}
\label{RFV}
\begin{aligned}
\overrightarrow{RVF}_{i, j} = [ZOTSim_{i, j}, \big| \|\overrightarrow{s}_i\| - \|\overrightarrow{s}_j\| \big|
, D_{i, j}]^{T}
\end{aligned}
\end{equation}
In \ref{RFV}, the second and third elements represent the relative speed and distance between vehicle $i$ and CH $j$, where $j$ is an element of the set $J$, representing the set of CHs such that $D_{i, j} \leq \min(TR_i, TR_j)$. Additionally, $ZOTSim_{i, j}$ is the zone-based overall trajectory similarity (ZOTSim) between $i$ and $j$. It is based on the cosine similarity of two vectors that indicate the general direction from the previous zones and current locations, defined as follows:
\begin{equation}
\label{sim}
\begin{aligned}
ZOTSim_{i, j} = cos^{-1}(\phi_{i, j})
\end{aligned}
\end{equation}
where, 
\begin{equation}
\label{phi}
\begin{aligned}
\phi_{i, j} = \frac{\overrightarrow{O}_i.\overrightarrow{O}_{j}}
{||\overrightarrow{O}_i||||\overrightarrow{O}_{j}||}
\end{aligned}
\end{equation}
Here, $\overrightarrow{O}_i$ represents a vector originating from the previous zone through which vehicle $v$ passed, extending to its current position, effectively illustrating the overall direction. Similarly, $\overrightarrow{O}_j$ is defined for the relevant CH. Considering a proper area for zones results in better trajectory and movement modeling of DRUs and vehicles. In this study, we encounter a trade-off when reducing zone size due to limited and brief TRs. There is a tendency to favor the direction closer to the current one instead of aligning with the desired overall direction, which will be automatically satisfied by $\overrightarrow{O}$.

Moreover, to calculate the CHEC for vehicle $i$ with respect to $ch$, it is essential to normalize each element of the variable $RVF_{i, j}$ using linear normalization as follows:
\begin{equation}
\label{normalized_RVF}
\begin{aligned}
\overrightarrow{RVF}_{norm, i, j} = N \odot \overrightarrow{RVF}_{i, j}
\end{aligned}
\end{equation}
where $N$ is an array defined as shown below:
\begin{equation}
\label{N}
\begin{aligned}
N = \left[\frac{1}{2\pi}, \frac{1}{\max(||\overrightarrow{s}_i||,||\overrightarrow{s}_j||)}, \frac{1}{\min(TR_i, TR_{j})}\right]
\end{aligned}
\end{equation}
It should be noted that since the minimum value for all elements in $\overrightarrow{RVF}_{i, j}$ is 0, (\ref{normalized_RVF}) provides linear normalization. However, to avoid the undefined form $\big| |\overrightarrow{s}_i| - |\overrightarrow{s}_j| \big|/0$, denominator regularization by adding $\epsilon$ should be applied. Therefore, considering all the CHs within $min\{TR_i, TR_j\}$, $v_i$ determines $RVF_i$ as follows:
\begin{equation}
\label{mat_RVF}
\begin{aligned}
RVF_i = &\begin{bmatrix}
    \overrightarrow{RFV}_{norm,i, j_1} & \ldots & \overrightarrow{RFV}_{norm,i, j_n}
\end{bmatrix} 
\end{aligned}
\end{equation}

Then, vehicle $v$ calculates the CHEC for all the mentioned CHs as follows:
\begin{equation}
\label{CHEC}
\begin{aligned}
CHEC_i = w^T(RVF_i)
\end{aligned}
\end{equation}
In \ref{CHEC}, $w$ is an array that assigns weights to $\overrightarrow{RVF}_{i, ch}$ where,
\begin{align}
\label{w_condition_2}
0 \leq w_l \leq 1, \\
\label{w_condition_1}
w_1 + w_2 + w_3 = 1
\end{align}
In the long run, $v_i$ identifies the most appropriate CH among the candidates by selecting the one that results in the lowest CHEC, as demonstrated below:
\begin{equation}
\label{CH_argmin}
\begin{aligned}
v_{ch, i} = \argmin_{j} \textit{CHEC}_i
\end{aligned}
\end{equation}
\begin{algorithm}[H]
\caption{Proposed CH Selection Strategy} 
\label{CH_selection}
\begin{algorithmic}[1]
\STATE \textit{$v_{dACK} \gets  \emptyset$, $v_{chACK} \gets \emptyset$, $v_{och} \gets \emptyset$,\\ $v_{ov} \gets \emptyset$, $v_{cmACK} \gets \emptyset$, and $v_{saACK} \gets \emptyset$}
\FOR{\textit{timing $=0$} \textbf{to} \textit{$\lambda\tau$} \textbf{by} \textit{$\tau$}}
    \STATE \textit{send beacons periodically}
    \STATE \textit{update $v_{dACK}$, $v_{chACK}$, $v_{cmACK}$, and $v_{saACK}$}
    \STATE $v_{ov} \gets v_{cmTR} \cup v_{saTR}$
    \IF{$v_{chACK} \cup v_{dACK} \neq \emptyset$}
        \STATE \textit{update links related to $v_{cmACK}$, and $v_{saACK}$}
        \STATE \textit{favoring $v_{dACK}$, generate $RVF$}
        \STATE \textit{calculate $CHEC$}
        \WHILE{($v_{ch} \neq \textbf{None}$) \OR ($CHEC \neq \emptyset$)}
            \STATE \textit{select CH using Eq. \ref{CH_argmin}}
            \IF{\textit{CH agrees}}
                \STATE $v_{ch} \gets$ \textit{CH}
                \STATE $v_{och} \gets v_{dACK} \cup v_{chACK} \setminus \{v_{ch}\}$
            \ELSE
                \STATE $CHEC \gets CHEC \setminus \{CH\}$
            \ENDIF
        \ENDWHILE
    \ENDIF
    \IF{($v_{ch}$ \textbf{is None}) \textbf{and} ($v_{saACK} \neq \emptyset$)}
        \STATE \textit{run Algorithm 3}
        \STATE \textbf{break}
    \ENDIF
\ENDFOR
\IF{($v_{ch}$ \textbf{is None}) \textbf{and} ($v_{saACK} = \emptyset$)}
    \STATE $ch \gets 1$
\ENDIF

\end{algorithmic}
\end{algorithm}

In the proposed approach, DRUs are prioritized as CHs due to their superior ability to meet the scalability requirements of VANETs. DRUs mainly operate along major roads, reducing the frequency of turns and detours. In contrast, other vehicles often take alternative routes to avoid traffic, resulting in more frequent diversions and exits. Additionally, DRUs are less likely to have significantly different $\overrightarrow{O}$ after changing zones. Moreover, DRUs rarely shut down or stop participating in VANETs. Therefore, if DRUs are in $TR_i$, the $RVF_i$ includes only them. Notably, after $\lambda\tau$ milliseconds, if $v_i$ cannot find a suitable cluster to join or form a cluster with other SAVs, it will set its $ch_i$ to 1. Here, $\tau$ is the period during which $v_i$ updates $v_{dACK, i}$, $v_{chACK, i}$, $v_{cmACK, i}$, and $v_{saACK, i}$ based on the received ACKs. Also, $\lambda \in \mathbb{N}$ limits the number of updates. Algorithm \ref{CH_selection} outlines the proposed decentralized SAVs' CH selection approach when they receive ACKs from CHs and DRUs in response to their beacons. In this algorithm, $v_{och, i}$ and $v_{ov, i}$ store other CHs, including DRUs, and other vehicles, both CMs and SAVs, within each other's TR, providing insights into packet routing options. Specifically, $v_{och, i}$ and $v_{ov, i}$ identify CMs that can act as gateways between two clusters. It should be noted that gateways would play a key role in the VANET expansion, reducing the number of connected components in favor of improving routing protocols.

\subsection{SAVs' Clusters Formation}
In this section, we explain the cluster formation of SAVs and the process for selecting a CH among them. According to Algorithm \ref{CH_selection}, if no neighboring clusters or CHs are discovered within $\lambda\tau$, each SAV announces itself as a potential CH (PCH) if other SAVs are nearby. They then disseminate this information along with the count of received ACKs from the SAVs that transmitted them. Since all SAVs follow the same approach, they update their PCH status by counting the PCHs of other SAVs within their TR. Each SAV adds the received PCHs to a set, neglecting PCH IDs not within its TR. Using \ref{normalized_RVF} and \ref{features}-\ref{CH_argmin}, CHs are determined, and clusters are formed.

If none of the received PCHs are within an SAV's TR, the SAV changes its $ch$ flag to 1. Additionally, if only two SAVs are within each other's TR, they both change their $ch$ flag to 1. This ensures that in zones with sparse traffic, more CH options are available, leading to increased stability in clusters and the network. Algorithm \ref{SAV_cluster} outlines this strategy. The cluster formation and CH selection process is designed to enhance cluster stability by unconsciously selecting the vehicle positioned centrally among others.
\begin{algorithm}[H]
\caption{SAVs Cluster Formation Strategy} 
\label{SAV_cluster}
\begin{algorithmic}[1]
\STATE \textit{$PCH \gets \emptyset$}
\STATE \textit{$v_{PCH} \gets id$}
\STATE \textit{share $n(v_{saACK})$ and $v_{PCH}$}
\STATE \textit{receive updated $v_{PCH, k}$, where $k \in v_{saACK}$}
\FOR{k in $v_{saACK}$}
    \STATE $temp \gets v_{PCH, k}$
    \IF{$n(v_{saACK, temp})\geq n(v_{saACK})$}
        \STATE $v_{PCH} \gets temp$
        \STATE \textit{$PCH \gets PCH \cup \{temp\}$}
    \ENDIF
\ENDFOR
\IF{$PCH \neq \emptyset$}
    \STATE \textit{calculate CHEC on PCH}
    \WHILE{($v_{ch} \neq \textbf{None}$)}
        \STATE $CH \gets$ \textit{select CH using Eq. \ref{CH_argmin}}
        \IF{\textit{(CH agrees)} $\&$ \textit{(CH is not a CM)}}
            \STATE \textit{$v_{ch} \gets CH$}
        \ELSE
            \STATE $CHEC \gets CHEC \setminus \{CH\}$
            \IF{$CHEC = \emptyset$}
                \STATE \textit{return to Algorithm 2}
            \ENDIF
        \ENDIF
    \ENDWHILE
\ENDIF
\end{algorithmic}
\end{algorithm}
\subsection{Cluster maintenance}
In the proposed approach, if a CH that is not a bus has no members, it will change its $ch$ flag to 0 and become an SAV as soon as it changes its zone. This eliminates the need for merging clusters and their associated complexities. Consequently, unnecessary CHs will become SAVs, while necessary CHs will maintain their status in the area. Thus, the number of CHs is automatically controlled by ensuring a diversity of directions to support new SAVs in selecting a CH.

In the event of a disconnection between SAVs, CMs, and CHs, the connection status will be updated after $\tau$ milliseconds to confirm the loss of connection. Therefore, each CM that loses its connection with its CH after periodic checks will employ Algorithms \ref{CH_selection} and \ref{SAV_cluster} after the specified time. Additionally, CHs periodically check their connections with each other.

\section{VANETs Clustering Evaluation Criteria}
This paper introduces a centralized evaluation factor called the VANETs Clustering Stability Metric (VCSM) to validate the proposed clustering algorithm's effectiveness. Evaluating clustering algorithms based on routing messages is not always appropriate because:
\begin{itemize}
    \item While enhancing cluster stability can improve message delivery and routing protocols in VANETs, the network's dynamic nature and the nodes' intermittent behavior introduce uncertainties. Consequently, there is no guarantee that clusters will remain connected to successfully pass packets, regardless of whether they take the minimum hop route.
    \item Cluster stability primarily enhances the routing protocol's effectiveness by more reliably coordinating the destination's CH based on the IP addressing structure. 
    \item This is particularly evident in this context, as once packets reach the destination zone (typically near the receiver), they are often broadcast to reach the receiver, independent of the clustering algorithm.
\end{itemize}

In this context, besides considering the number of connected components to reflect routing capability, VCSM is defined by the frequency of cluster changes by a vehicle and the duration they remain in the clusters they join. Therefore, it is defined as follows:
\begin{equation}
\label{VCSM}
\begin{aligned}
VCSM = \frac{1}{n_{vm}}\Gamma\Theta
\end{aligned}
\end{equation}
In \ref{VCSM}, $n_{vm}$ is the total number of vehicles coming and leaving the understudies area and joining a cluster as a CM at least once. Additionally, $\Gamma$ is a coefficient array for the purpose of normalization defined as:
\begin{equation}
\label{Gamma}
\begin{aligned}
\Gamma = &\begin{bmatrix}
    &\frac{1}{\gamma_1T_1} & \cdots & \frac{1}{\gamma_{n_{vm}}T_{n_{vm}}} 
\end{bmatrix} 
\end{aligned}
\end{equation}
where $\gamma_i$ and $T_i$ represent the number of the clusters that vehicle $i$ has joined and the total amount of time of its attendance in the studied area, respectively. Additionally, $\Theta$ is a sparse matrix, and its elements show how long each vehicle remained in one cluster, as follows:
\begin{equation}
\label{Theta}
\begin{aligned}
\Theta = &\begin{bmatrix}
    &t_{1,1}   & \cdots & \cdots \\
    &\vdots & \ddots & \vdots \\
    &t_{n_{vm}, 1} & \cdots & \cdots
\end{bmatrix} 
\end{aligned}
\end{equation}
It should be noted that the size of $\Theta$ is $(n_v, \max(\gamma_i))$. Due to the varying number of clusters each vehicle might join, each row in $\Theta$, corresponding to a vehicle, may contain different numbers of zeros.

In an ideal scenario, no vehicles would act as CHs, and all would be CMs of DRUs connected to each other. They would remain as CMs of a single cluster until they leave the area or are turned off, resulting in a VCSM of 1. However, metropolitan areas often lack sufficient DRUs, necessitating other vehicles to form clusters and create VANETs. In this context, VCSM remains 1 if, in addition to the DRUs, CHs maintain their status from the beginning and their CMs do not leave their clusters. Nevertheless, CMs do not always follow CHs and DRUs due to changing directions and trajectories, as well as limited TR, altering the ideal scenario. Here, $\gamma_i$ acts as a penalty factor in VCSM, reducing stability based on the number of clusters a vehicle $i$ joins. Therefore, as discussed in Section \ref{sec: area-preparation}, $w$ is another piece of data that DRUs and CHs share with CMs and SAVs according to varying traffic conditions, typically updated every few hours.

The significance of VCSM is clear when considering that multiple SAVs may unintentionally be disconnected from the network, which does not affect the evaluation. The key factor influencing VCSM is the intentional decision-making of SAVs in selecting a CH along with TR.
\section{Experimental Results}
In this section, we assess the SMZCA introduced in this paper. We offer insights into the implementation framework, tools, and input data details. We then elaborate on the parameters used in our evaluation process. Following this, we present the output results from the case studies and comparisons. Finally, we discuss the results, emphasizing the effectiveness of the proposed contributions.
\subsection{Experimental setting}
\begin{figure}[htp]
\centering
\includegraphics[width=3.5in, height=3.25in]{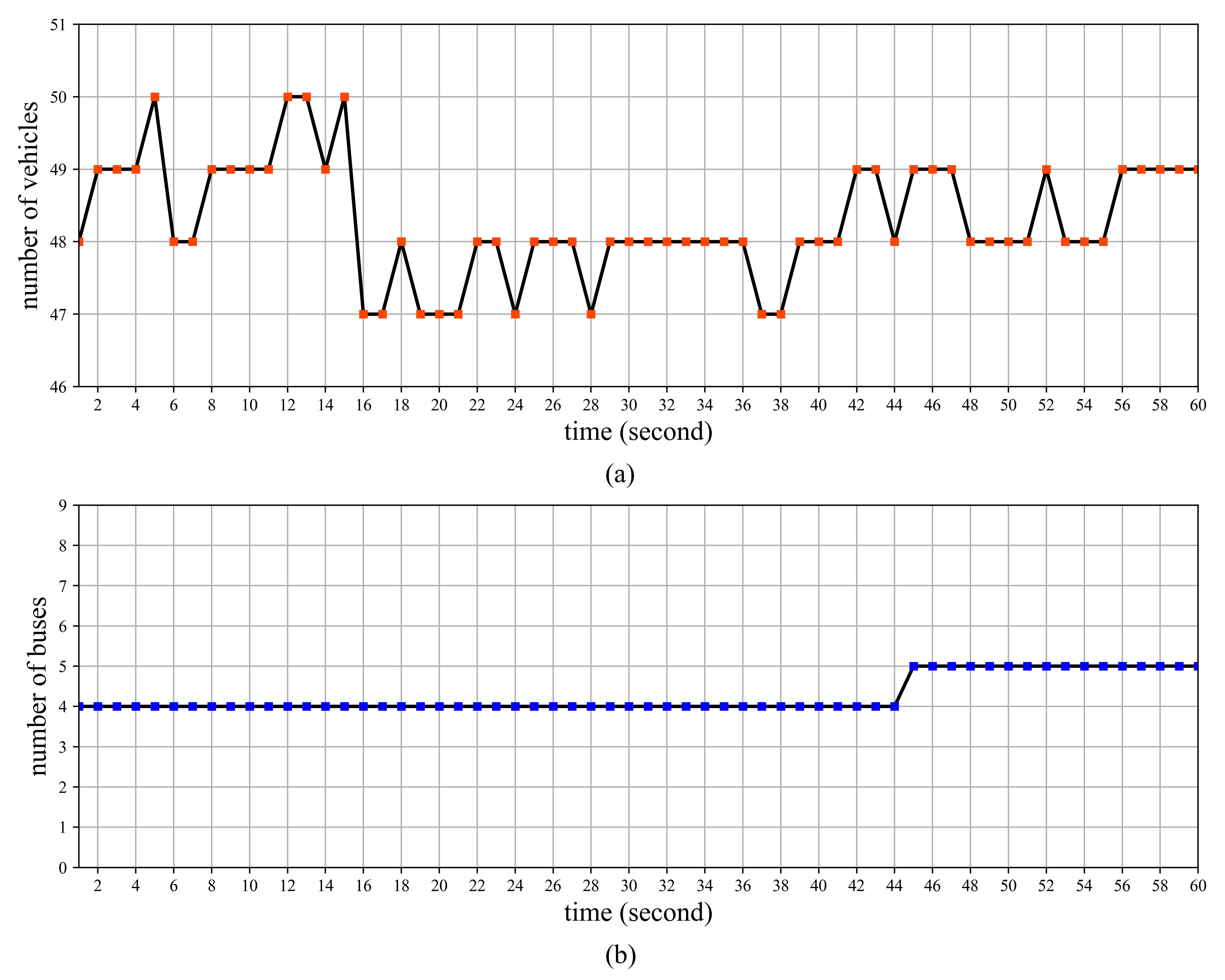} 
\caption{Number of nodes within the understudied area over the $60s$ period. (a) number of vehicles. (b) number of buses}
\label{nbuses_nvehs}
\end{figure}

We implemented the proposed SMZCA in Python, employing datasets generated in SUMO for $\approx 6.5km^2$ area, after padding, of the Greater Toronto Area (GTA), ON, Canada. To make the conditions more challenging, we did the following:
\begin{itemize}
    \item Even though metropolitan areas typically experience high vehicular congestion due to a significant number of vehicles, exclusive of public transportation, bicycles, and emergency vehicles, leading to dense traffic jams, which positively affect VCSM for small regions, a very sparse traffic data with only up to fifty cars and five buses in the understudied area is considered for the proposed case studies in the next subsection.
    \item As discussed in \ref{sec: proposed SMZCA}, only public buses are employed as DRUs.
    \item Different TRs are applied to vehicles all using the same TR.
    \item Even the outliers regarding VCSM obtained from some of the $w$'s are counted too.
\end{itemize}
As previously discussed, adjusting each zone's area should be tailored to the most common TR of vehicles. According to the Dedicated Short-Range Communication (DSRC) standardized communication protocol, as shown in Table \ref{TR-alpha}, recommended values for $\alpha$ regarding normal metropolitan areas' traffic are presented. In practice, vehicles have different maximum TRs due to their antenna characteristics, which depend on factors such as antenna type, design, and energy efficiency trade-offs related to vehicle size and design. Therefore, in this paper, three different TRs are assigned separately as the common TR in the area (applied to all vehicles) to highlight the significant performance of the proposed method. While some studies have explored DSRC's potential for operation up to a range of $1000m$~\cite{Rawat-Shetty, Taha-Hasan}, our study adopts a conservative approach, using TRs of $100m$, $200m$, and $300m$ for vehicles, while the TR for DRUs is set at $800m$ for all cases. Additionally, all possible weights ranging from $0$ to $1$ with a step size of $0.1$ are assigned to $w$ to calculate CHEC. These weights are consistent for all vehicles. To ensure mature data, a $60$-second consecutive time period between the $1500^{th}$ and $2000^{th}$ second is selected from the data generated by SUMO. Table~\ref{other_parameters} shows other assumptions and parameters used in this study. Furthermore, the number of CMs that a bus and other CHs can have is limited to $30$ and $20$, respectively. However, the algorithm is flexible, and these limitations are not strict, as traffic situations can vary in practice, depending on the application of VANETs. Moreover, Fig~\ref{nbuses_nvehs} shows the number of buses and vehicles in the area for the proposed case studies.

\begin{table}[htp]
    \centering
    \caption{Suggested values for $\alpha$ to Adjust Zone Sizes Based on TR}
    \label{TR-alpha}
    \begin{tabular}{|c|c|}
        \hline
        TR & $\alpha$ \\
        \hline
        $100m-300m$ \cite{Rezgui-Charkaui, Hafeez-Anpalagan} & $0.5$ \\
        \hline
        $400m-500m$ \cite{Sommer-Eckholff, Ma-Zhang} & $0.8$ \\
        \hline
        $600m-1000m$ \cite{Zeng-Yu, Khan-Koubaa, Rawat-Shetty, Taha-Hasan} & $1$ \\
        \hline
    \end{tabular}
\end{table}

\begin{table}[htp]
    \centering
    \caption{Parameters Employed in SMZCA Implementation}
    \label{other_parameters}
    \begin{tabular}{|c|c|}
        \hline
        Parameters & Values \\
        \hline
        $\lambda$& $4$ \\
        \hline
        $\tau$&$1000$ \textit{milliseconds}\\
        \hline
        \emph{start-time}& $1601$ \\
        \hline
        \emph{end-time}& $1660$ \\
        \hline
        $lat^{min}$ \emph{(after padding)}&$43.860130$\\
        \hline
        $long^{min}$ \emph{(after padding)}&$-79.462871$\\
        \hline
        $lat^{max}$ \emph{(after padding)}&$43.8795$\\
        \hline
        $long^{max}$ \emph{(after padding)}&$-79.432551$\\
        \hline
    \end{tabular}
\end{table}

\subsection{Case studies}
In this paper, we study the performance of the proposed SMZCA using the following two scenarios:
\begin{enumerate}
    \item Only DRUs are in the area being used
    \item Both DRUs and two RSUs are in the area. The static RSUs are at two bus stations at two main intersections, which are not in the residential areas.
\end{enumerate}
For both scenarios, all the considered TRs and weights are applied separately over a consecutive $60$-second period. It is important to note that the TR for DRUs and RSUs is set higher and fixed at $800m$, although this does not impact the clustering process.

\subsection{Numerical results and discussion}
\begin{figure}[htp]
\centering
\includegraphics[height=2in]{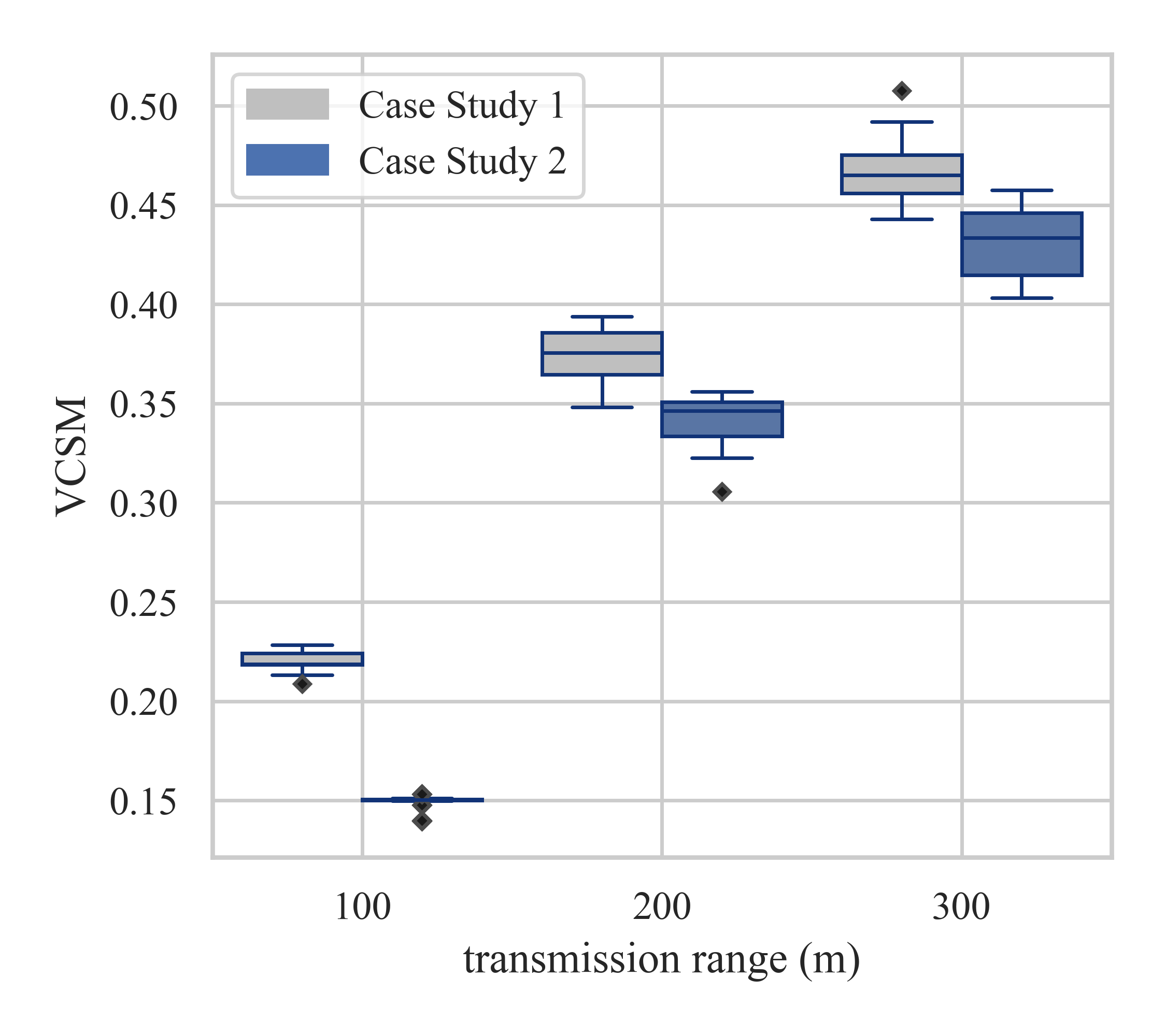} 
\caption{Distribution of the calculated VCSM across diverse $w$ values for all evaluated TRs in both case studies}
\label{eval_boxplots}
\end{figure}

Figure \ref{eval_boxplots} shows the distribution of obtained VCSM for the proposed clustering algorithm's performance in both case studies, considering all possible values for $w$s with a step-size of $0.1$. As observed, the presence of RSUs does not enhance cluster stability for any TRs; in fact, it reduces it. However, the significant performance of SMZCA results in the highest VCSM for all TRs in both case studies. SMZCA's ability to improve cluster stability is evident, as the lowest VCSMs are associated with $w$s that assign low weights to $ZOTSim$. Therefore, regardless of RSU usage, implementing SMZCA significantly enhances cluster stability. Tables \ref{TR-omega-eval-cs1} and \ref{TR-omega-eval-cs2} show the $w$s for different TRs in the first and second case studies, respectively, leading to the best and worst VCSMs. It is worth noting that, regardless of the number of vehicles in the area, the finite number of CHs and DRUs within an SAV's TR means that similar weights may result in the same VCSM.

\begin{table}[htp]
    \centering
    \caption{$w$s leading to Maximum and Minimum VCSM for Case Study 1}
    \label{TR-omega-eval-cs1}
    \begin{tabular}{|c|c|c|c|}
        \hline
        TR & $w\Rightarrow\min(VCSM)$ & $w\Rightarrow\max(VCSM)$ & Improvement \\
        \hline
        $100$  & $[0.0, 0.8, 0.2]$  & $[0.5, 0.5, 0.0]$ &  $\approx 15\%$\\
        \hline
        $200$  & $[0.1, 0.1, 0.8]$  & $[0.5, 0.2, 0.3]$ & $\approx 14.7\%$\\
        \hline
        $300$  & $[0.0, 0.8, 0.2]$  & $[0.9, 0.0, 0.1]$ & $\approx 15.9\%$ \\
        \hline
    \end{tabular}
\end{table}

\begin{table}[htp]
    \centering
    \caption{$w$s leading to Maximum and Minimum VCSM for Case Study 2}
    \label{TR-omega-eval-cs2}
        \begin{tabular}{|c|c|c|c|}
        \hline
        TR & $w\Rightarrow\min(VCSM)$ & $w\Rightarrow\max(VCSM)$ & Improvement \\
        \hline
        $100$  & $[0.0, 0.0, 1.0]$  & $[0.9, 0.1, 0.0]$ &  $\approx 10\%$\\
        \hline
        $200$  & $[0.0, 1.0, 0.0]$  & $[0.6, 0.2, 0.2]$ & $\approx 16\%$\\
        \hline
        $300$  & $[0.2, 0.7, 0.1]$  & $[0.7, 0.0, 0.3]$ & $\approx 13\%$ \\
        \hline
    \end{tabular}
\end{table}
There are several crucial aspects to consider. Firstly, while many studies emphasize relative speed, distance, or other relevant features as key factors for determining CHs, our results underscore the capability of the proposed $ZOTSim_{i, j}$ in comparison. Specifically, across both case studies and for all TRs, the VCSMs lower than the medians are associated with $w$s that have dominant weights assigned to these conventional factors. This does not diminish the importance of these metrics but rather highlights the unique capability of $ZOTSim_{i, j}$ in achieving optimal CH selection. Secondly, it is evident that for all TRs, the results of the first case study, which fall between the lower and upper quartiles, are higher than those in the second case study. To elaborate, in practice, selecting $w$ would be significantly more sensitive if RSUs are deployed. This implies that the presence of RSUs as participants in clusters generally reduces the stability of VANETs in clusters.

\begin{figure}[htp]
\centering
\includegraphics[width=3.5in, height=2.5in]{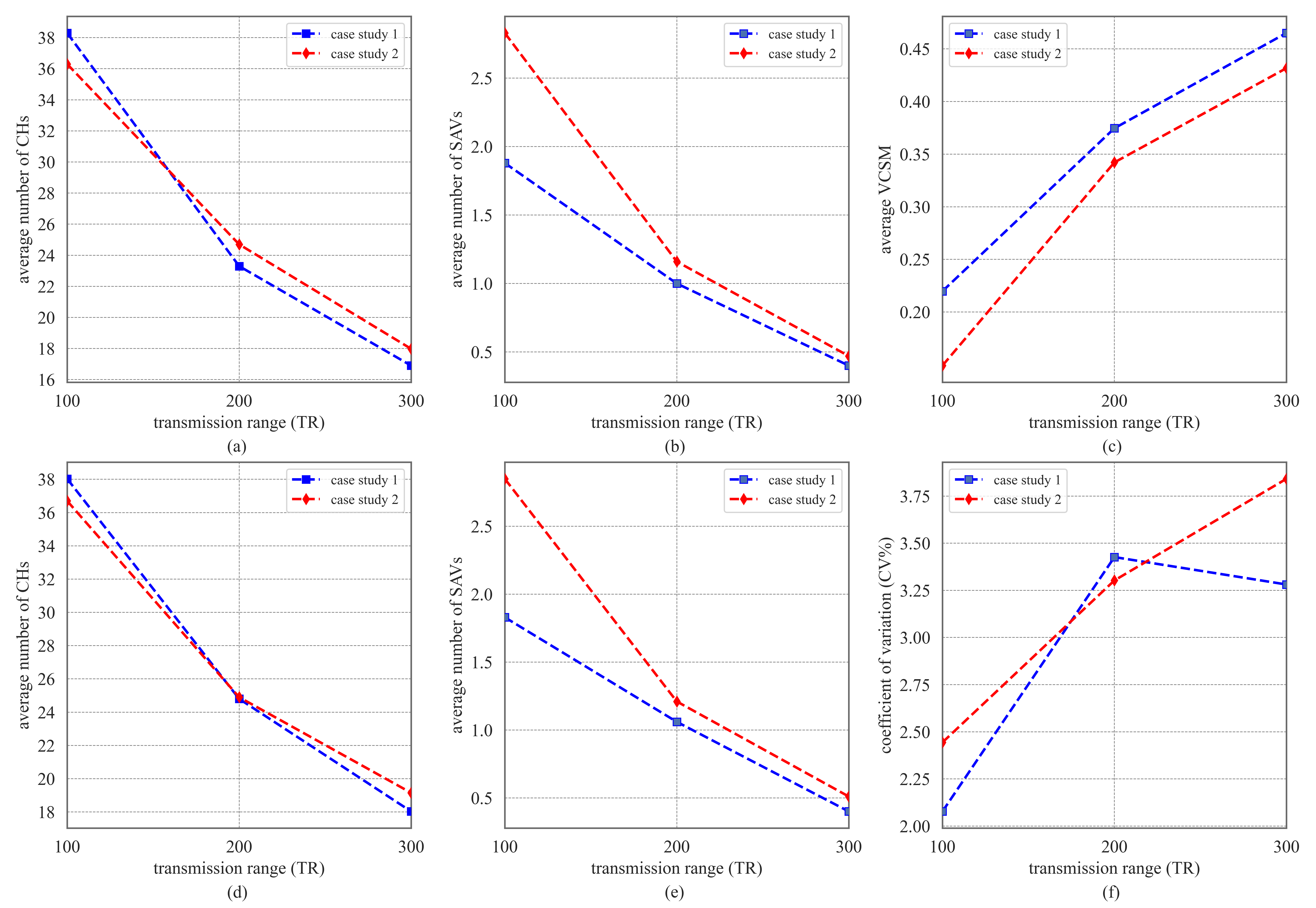} 
\caption{SMZCA performance for different TRs in both case studies. (a) average number of CHs for the best scenario regarding VCSM. (b) average number of SAVs for the best scenario regarding VCSM. (c) average VCSM of all weights. (d) the average number of CHs during the $60s$ period for the worst scenario regarding VCSM. (e) average number of SAVs for the worst scenario regarding VCSM. (f) $CV\%$ of VCSM for all weights.}
\label{six_plots}
\end{figure}

Figure \ref{six_plots} provides a detailed comparison of the two case studies by examining the worst and best $w$s leading to VCSMs. This figure highlights the superior performance of the first case study, leveraging the average number of CHs, average number of SAVs, and average VCSM. When $TR=100m$, although the average number of CHs is slightly lower in the second case study, the better performance of the algorithm in the first case study is evident when considering Figures \ref{six_plots}(b), \ref{six_plots}(c), and \ref{six_plots}(d), which reflect the average number of SAVs and average VCSM. Conversely, for $TR \in [200m, 300m]$, the first case study consistently yields better results for both the worst and best $w$s. With the presence of RSUs, vehicles have more options for selecting CHs, but they are also more likely to become SAVs quickly, form their clusters, or announce themselves as CHs after a specific time. Additionally, the lower average VCSMs for all TRs in the second case study align with its increased vulnerability due to having more options for $w$.

Finally, to compare the variations in computed VCSMs from two datasets—excluding and including RSUs for the first and second case studies—the coefficient of variation ($CV\%$) is used as follows: 
\begin{equation}
\label{cv}
\begin{aligned}
CV\% = \frac{S_{VCSM}}{\overline{VCSM}}\times100
\end{aligned}
\end{equation}
where $S_{VCSM}$ is the standard deviation, and $\overline{VCSM}$ is the mean value of obtained VCSMs for the selected weights. As lower $CV\%$ implies lower risk, it can be seen when $TR=200m$, case study one has higher risk, although not significant for $CV\%$ considering the efficiency of the proposed method. Here the argument against relative speed and distance factors can be justified once more since the outliers cause a higher risk when RSUs are not deployed (see Fig. \ref{eval_boxplots}, and Table \ref{TR-omega-eval-cs1}).

To further justify the performance of the SMZCA, a comprehensive comparison is made with a Befit factor-based clustering algorithm proposed in \cite{arkian-Atani} and a dynamic single-hop clustering algorithm (DSCA) built on top of it, presented in \cite{Katiyar-Gupta}. These are two single-hop clustering algorithms in VANETs. Due to the limited availability of established clustering algorithms for metropolitan areas, particularly for single-hop clustering, SMZCA is compared with highway-based clustering algorithms as a baseline. Key performance metrics, such as cluster stability and overhead, are relevant across both highway and metropolitan environments, supporting the validity of this comparison. To ensure fairness, the DRUs and RSUs are removed from the datasets, and three different datasets with varying sparsity are employed. This comparison highlights the adaptability and effectiveness of our algorithm in handling the complex dynamics of urban traffic, demonstrating its potential value across diverse environments.

\begin{figure}[htp]
\centering
\includegraphics[width=3.5in, height=4in]{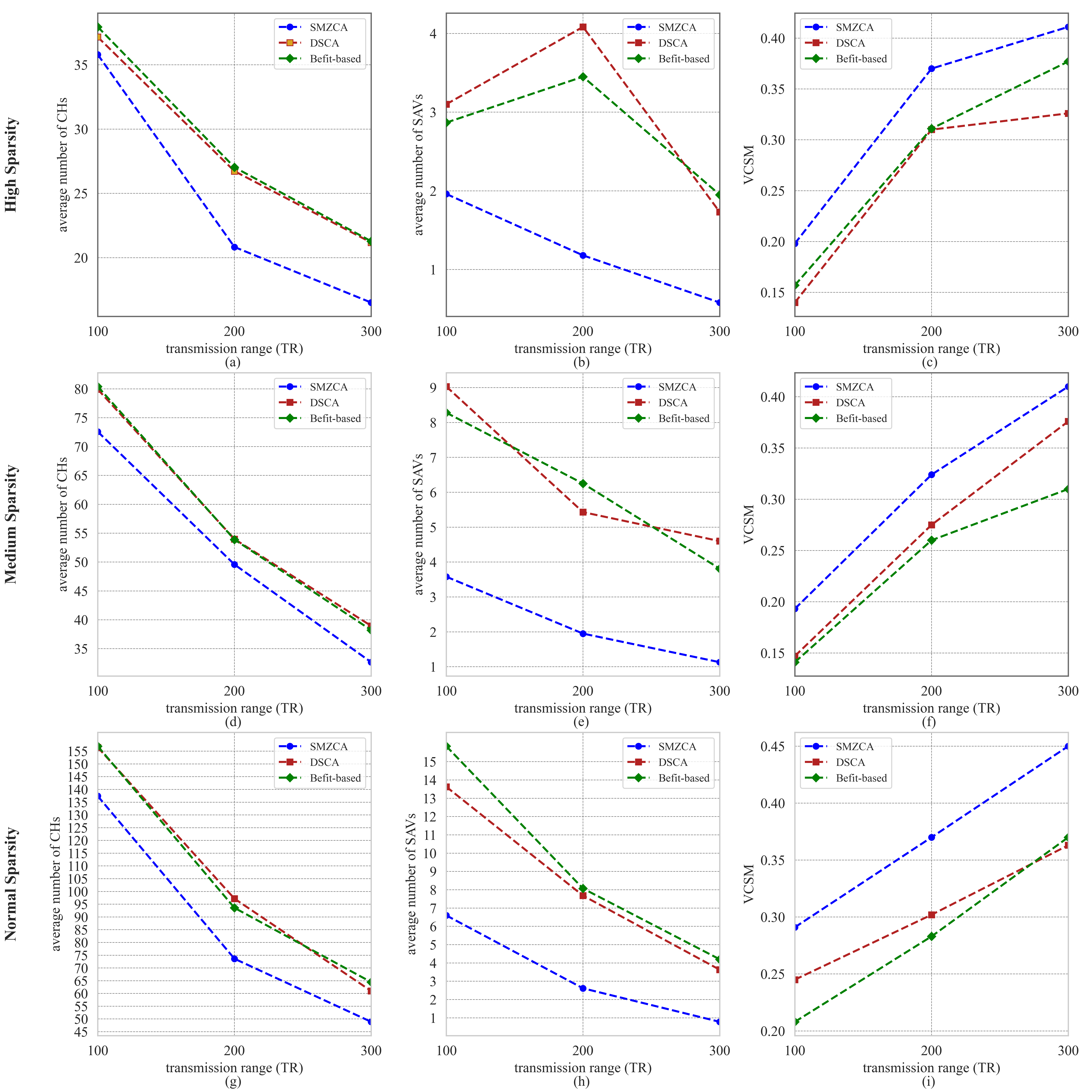} 
\caption{Comparison of SMZCA with DSCA and Befit-based algorithms through different vehicle sparsity and TRs. (a) the average number of CHs using a dataset with high sparsity. (b) average number of SAVs using a dataset with high sparsity. (c) obtained VCSM for each algorithm using a dataset with high sparsity. (d) the average number of CHs using a dataset with medium sparsity. (e) average number of SAVs using a dataset with medium sparsity. (f) obtained VCSM for each algorithm using a dataset with medium sparsity. (g) the average number of CHs using a dataset with normal sparsity. (h) average number of SAVs using a dataset with normal sparsity. (i) obtained VCSM for each algorithm using a dataset with normal sparsity.}
\label{comparison_plots}
\end{figure}

Figure \ref{comparison_plots} presents a comprehensive performance evaluation of SMZCA in comparison with DSCA and the Befit-based algorithm across different vehicle sparsity levels and TRs. The results indicate that the proposed SMZCA algorithm consistently outperforms the alternatives in terms of the average number of CHs, SAVs, and VCSM. A key observation is the substantial reduction in the number of SAVs, with SMZCA demonstrating even more than $50\%$ decrease in most cases. This reduction is particularly important as fewer SAVs indicate a more efficient clustering mechanism, leading to improved connectivity and communication reliability. Furthermore, it is noteworthy that SMZCA achieves such performance with a lower number of CHs, which translates to reduced overhead and better resource utilization. Another critical aspect to highlight is that, despite the differences in CH and SAV counts, all algorithms maintain the same average number of connected components for each TR and sparsity level. This consistency validates SMZCA’s ability to integrate seamlessly into routing protocols without compromising network connectivity. Additionally, Table \ref{algorithms_comparison} provides a detailed comparison of the overhead and the information required for vehicles to operate under each algorithm’s protocol, further demonstrating the efficiency of SMZCA. Beyond these comparisons, it is essential to mention that the dataset used for the high vehicle sparsity scenario is the same as that utilized in case studies where dedicated RSUs and DRUs are absent. Despite this challenging condition, SMZCA maintains a satisfactory VCSM across all TRs, indicating its robustness in managing cluster stability solely through vehicle-based communication. These findings emphasize the effectiveness of SMZCA in improving clustering performance, reducing isolation, and ensuring stability in VANETs, particularly in metropolitan environments.

\begin{table}[htp]
    \centering
    \caption{Information Required by SMZCA, DSCA, and Befit Algorithms}
    \label{algorithms_comparison}
    \begin{tabular}{|c|c|c|c|}
        \hline
        \multirow{2}{*}{Information} & \multicolumn{3}{c|}{Algorithms} \\
        \cline{2-4}
         & SMZCA & DSCA & Befit-based \\
        \hline
        locations & \checkmark & \checkmark & \checkmark \\
        \hline
        number of neighbours & \checkmark & \checkmark & \checkmark \\
        \hline
        intersections' locations & --- & \checkmark & \checkmark \\
        \hline
        speed & \checkmark & \checkmark & \checkmark \\
        \hline
        time duration of covering distances & --- & \checkmark & \checkmark \\
        \hline
    \end{tabular}
\end{table}

\section{Conclusion}
This paper introduces a clustering algorithm tailored for metropolitan areas by excluding RSUs and utilizing area zoning to enhance cluster stability. Additionally, a centralized measurement is proposed to evaluate cluster stability based on the decision-making process of vehicles in selecting a CH and forming clusters. To demonstrate the effectiveness of the proposed algorithm, challenging conditions such as extremely sparse vehicle density, a low number of buses, and public transportation limitations are considered. The analysis of two case studies and a comprehensive comparison with two existing single-hop clustering algorithms validate the competency of the proposed SMZCA in improving cluster stability. Furthermore, the results indicate that the presence of RSUs leads to more frequent cluster changes, reducing their effectiveness in high-mobility VANETs. This also results in the possibility of having more SAVs as well as CHs from vehicles, increasing the number of hops in the routing process and packet losses. However, the role of RSUs in expanding VANETs cannot be overlooked, and the trade-off with relevant investments in metropolitan areas warrants further study. Additionally, proper weight selection is crucial for better cluster stability. The proposed $ZOTsim$ factor has shown significant performance compared to conventional metrics like relative speed and distance. Following this study, the next steps would be to fill the gaps related to multi-hop clustering and design appropriate routing algorithms for both frameworks.

\bibliographystyle{IEEEtran}
\bibliography{references}

\begin{thebibliography}{10}
\providecommand{\url}[1]{#1}
\csname url@samestyle\endcsname
\providecommand{\newblock}{\relax}
\providecommand{\bibinfo}[2]{#2}
\providecommand{\BIBentrySTDinterwordspacing}{\spaceskip=0pt\relax}
\providecommand{\BIBentryALTinterwordstretchfactor}{4}
\providecommand{\BIBentryALTinterwordspacing}{\spaceskip=\fontdimen2\font plus
\BIBentryALTinterwordstretchfactor\fontdimen3\font minus \fontdimen4\font\relax}
\providecommand{\BIBforeignlanguage}[2]{{%
\expandafter\ifx\csname l@#1\endcsname\relax
\typeout{** WARNING: IEEEtran.bst: No hyphenation pattern has been}%
\typeout{** loaded for the language `#1'. Using the pattern for}%
\typeout{** the default language instead.}%
\else
\language=\csname l@#1\endcsname
\fi
#2}}
\providecommand{\BIBdecl}{\relax}
\BIBdecl

\bibitem{Kokare-Kakkar2021}
M.~B. Kokare and D.~Kakkar, ``A survey on clustering algorithms for cluster-head selection in {VANET},'' in \emph{2021 Second International Conference on Electronics and Sustainable Communication Systems (ICESC)}, 2021, pp. 992--996.

\bibitem{Sarkar-Chakrabarty2021}
R.~R. Sarkar, A.~Chakrabarty, and M.~Z. Rahman, ``{VANET} routing protocols in real-world mobility tracing,'' in \emph{2021 13th International Conference on Computational Intelligence and Communication Networks (CICN)}, 2021, pp. 96--101.

\bibitem{Zhang-Ge}
D.~Zhang, H.~Ge, T.~Zhang, Y.-Y. Cui, X.~Liu, and G.~Mao, ``New multi-hop clustering algorithm for vehicular ad hoc networks,'' \emph{IEEE Transactions on Intelligent Transportation Systems}, vol.~20, no.~4, pp. 1517--1530, 2019.

\bibitem{Bi-Shan}
Y.~Bi, H.~Shan, X.~S. Shen, N.~Wang, and H.~Zhao, ``A multi-hop broadcast protocol for emergency message dissemination in urban vehicular ad hoc networks,'' \emph{IEEE Transactions on Intelligent Transportation Systems}, vol.~17, no.~3, pp. 736--750, 2016.

\bibitem{Ren-Zhang}
M.~Ren, J.~Zhang, L.~Khoukhi, H.~Labiod, and V.~Vèque, ``A unified framework of clustering approach in vehicular ad hoc networks,'' \emph{IEEE Transactions on Intelligent Transportation Systems}, vol.~19, no.~5, pp. 1401--1414, 2018.

\bibitem{Abbas-Abdulsattar2022}
A.~H. Abbas, N.~F. Abdulsattar, H.~S. Mansour, M.~H. Mutar, M.~I. Habelalmateen, L.~Audah, N.~Alduais, and A.~Mohammed, ``A new hybrid approach cluster-heads election to reduce the number of clusters in {VANET}s,'' in \emph{2022 5th International Conference on Engineering Technology and its Applications (IICETA)}, 2022, pp. 436--440.

\bibitem{Saleem-Zhou2019}
M.~A. Saleem, S.~Zhou, A.~Sharif, T.~Saba, M.~A. Zia, A.~Javed, S.~Roy, and M.~Mittal, ``Expansion of cluster head stability using fuzzy in cognitive radio {CR-VANET},'' \emph{IEEE Access}, vol.~7, pp. 173\,185--173\,195, 2019.

\bibitem{Ferng-Abdullah2019}
H.-W. Ferng and M.~Abdullah, ``Mobility-based clustering with link quality estimation for urban {VANET}s,'' in \emph{2019 International Conference on Machine Learning and Cybernetics (ICMLC)}, 2019, pp. 1--7.

\bibitem{Liang-Wang}
B.~Liang, F.~Wang, and B.~Ran, ``Optimizing roadside unit deployment in vanets: A study on consideration of failure,'' \emph{IEEE Transactions on Intelligent Transportation Systems}, 2024.

\bibitem{Jain-Jeyakumar2016}
K.~Jain and A.~Jeyakumar, ``An {RSU} based approach: A solution to overcome major issues of routing in {VANET},'' in \emph{2016 International Conference on Communication and Signal Processing (ICCSP)}, 2016, pp. 1265--1269.

\bibitem{Yeung-Hui}
C.~Y. Yeung, L.~C.~K. Hui, T.~W. Chim, S.-M. Yiu, G.~Zeng, and J.~Chen, ``Anonymous counting problem in trust level warning system for {VANET},'' \emph{IEEE Transactions on Vehicular Technology}, vol.~68, no.~1, pp. 34--48, 2019.

\bibitem{Jalooli-Song}
A.~Jalooli, M.~Song, and X.~Xu, ``Delay efficient disconnected {RSU} placement algorithm for {VANET} safety applications,'' in \emph{2017 IEEE Wireless Communications and Networking Conference (WCNC)}, 2017, pp. 1--6.

\bibitem{Benkirane}
S.~Benkirane, ``Road safety against sybil attacks based on {RSU} collaboration in {VANET} environment,'' in \emph{Mobile, Secure, and Programmable Networking: 5th International Conference, MSPN 2019, Mohammedia, Morocco, April 23--24, 2019, Revised Selected Papers 5}.\hskip 1em plus 0.5em minus 0.4em\relax Springer, 2019, pp. 163--172.

\bibitem{Zhang-Li2021}
Z.~Zhang, C.~Li, L.~Yu, Y.~Zhao, and Y.~Li, ``A multi-objective roadside units deployment method in {VANET},'' in \emph{2021 IEEE International Conference on Smart Internet of Things (SmartIoT)}, 2021, pp. 390--394.

\bibitem{Kim-Velasco}
D.~Kim, Y.~Velasco, W.~Wang, R.~N. Uma, R.~Hussain, and S.~Lee, ``A new comprehensive {RSU} installation strategy for cost-efficient {VANET} deployment,'' \emph{IEEE Transactions on Vehicular Technology}, vol.~66, no.~5, pp. 4200--4211, 2017.

\bibitem{Jabbar-Trabelsi2022}
M.~K. Jabbar and H.~Trabelsi, ``A betweenness centrality based clustering in {VANET}s,'' in \emph{2022 15th International Conference on Security of Information and Networks (SIN)}, 2022, pp. 1--4.

\bibitem{arkian-Atani}
H.~R. Arkian, R.~E. Atani, A.~Pourkhalili, and S.~Kamali, ``Cluster-based traffic information generalization in vehicular ad-hoc networks,'' \emph{Vehicular communications}, vol.~1, no.~4, pp. 197--207, 2014.

\bibitem{Katiyar-Gupta}
A.~Katiyar, S.~K. Gupta, D.~Singh, and R.~S. Yadav, ``A dynamic single-hop clustering algorithm ({DSCA}) in {VANET},'' in \emph{2020 11th International Conference on Computing, Communication and Networking Technologies (ICCCNT)}, 2020, pp. 1--6.

\bibitem{Khan-Abolhasan2018}
A.~A. Khan, M.~Abolhasan, and W.~Ni, ``An evolutionary game theoretic approach for stable and optimized clustering in {VANET}s,'' \emph{IEEE Transactions on Vehicular Technology}, vol.~67, no.~5, pp. 4501--4513, 2018.

\bibitem{Affandi-Suardi2021}
A.~Affandi, D.~Suardi, E.~Setijadi, G.~Kusrahardjo, and Endroyono, ``Application of clustering method on vehicular ad-hoc network ({VANET}) on mobility of medical vehicles in urban environment,'' in \emph{2021 4th International Seminar on Research of Information Technology and Intelligent Systems (ISRITI)}, 2021, pp. 618--623.

\bibitem{Cheng-Yuan2020}
J.~Cheng, G.~Yuan, M.~Zhou, S.~Gao, Z.~Huang, and C.~Liu, ``A connectivity-prediction-based dynamic clustering model for {VANET} in an urban scene,'' \emph{IEEE Internet of Things Journal}, vol.~7, no.~9, pp. 8410--8418, 2020.

\bibitem{Alsuhli-Khattab2019}
G.~H. Alsuhli, A.~Khattab, Y.~A. Fahmy, and Y.~Massoud, ``Enhanced urban clustering in {VANET}s using online machine learning,'' in \emph{2019 IEEE International Conference on Vehicular Electronics and Safety (ICVES)}, 2019, pp. 1--6.

\bibitem{Shah-Habib2018}
Y.~A. Shah, H.~A. Habib, F.~Aadil, M.~F. Khan, M.~Maqsood, and T.~Nawaz, ``Camonet: Moth-flame optimization (mfo) based clustering algorithm for {VANET}s,'' \emph{IEEE Access}, vol.~6, pp. 48\,611--48\,624, 2018.

\bibitem{Tseng-Wu2020}
H.-W. Tseng, R.-Y. Wu, and C.-W. Lo, ``A stable clustering algorithm using the traffic regularity of buses in urban {VANET} scenarios,'' \emph{Wireless Networks}, vol.~26, pp. 2665--2679, 2020.

\bibitem{Abboud-Zhuang}
K.~Abboud and W.~Zhuang, ``Stochastic modeling of single-hop cluster stability in vehicular ad hoc networks,'' \emph{IEEE Transactions on Vehicular Technology}, vol.~65, no.~1, pp. 226--240, 2016.

\bibitem{Wang-Mitton2023}
B.~Wang, N.~Mitton, and J.~Zheng, ``Prezcast: A preferred-zone based broadcast protocol for urban areas of {VANET}s,'' in \emph{IEEE International Conference on Communications (ICC)}, 2023.

\bibitem{Lin-Kang2017}
D.~Lin, J.~Kang, A.~Squicciarini, Y.~Wu, S.~Gurung, and O.~Tonguz, ``Mozo: A moving zone based routing protocol using pure v2v communication in {VANET}s,'' \emph{IEEE Transactions on Mobile Computing}, vol.~16, no.~5, pp. 1357--1370, 2017.

\bibitem{Kitsis-Datta2018}
R.~Kitsis and S.~Datta, ``Layer 3 enhancements for vehicular ad hoc networks,'' in \emph{The 9th International Conference on Ambient Systems, Networks and Technologies {(ANT} 2018), May 8-11, 2018, Porto, Portugal}, vol. 130.\hskip 1em plus 0.5em minus 0.4em\relax Elsevier, 2018, pp. 628--635.

\bibitem{Rawat-Shetty}
D.~B. Rawat and S.~Shetty, ``Enhancing connectivity for spectrum-agile vehicular ad hoc networks in fading channels,'' in \emph{2014 IEEE Intelligent Vehicles Symposium Proceedings}.\hskip 1em plus 0.5em minus 0.4em\relax IEEE, 2014, pp. 957--962.

\bibitem{Taha-Hasan}
M.~M.~I. Taha and Y.~M.~Y. Hasan, ``{VANET}-dsrc protocol for reliable broadcasting of life safety messages,'' in \emph{2007 IEEE International Symposium on Signal Processing and Information Technology}, 2007, pp. 104--109.

\bibitem{Rezgui-Charkaui}
J.~Rezgui, S.~Cherkaoui, and O.~Chakroun, ``Deterministic access for dsrc/802.11p vehicular safety communication,'' in \emph{2011 7th International Wireless Communications and Mobile Computing Conference}, 2011, pp. 595--600.

\bibitem{Hafeez-Anpalagan}
K.~A. Hafeez, A.~Anpalagan, and L.~Zhao, ``Optimizing the control channel interval of the dsrc for vehicular safety applications,'' \emph{IEEE Transactions on Vehicular Technology}, vol.~65, no.~5, pp. 3377--3388, 2016.

\bibitem{Sommer-Eckholff}
C.~Sommer, D.~Eckhoff, and F.~Dressler, ``Ivc in cities: Signal attenuation by buildings and how parked cars can improve the situation,'' \emph{IEEE Transactions on Mobile Computing}, vol.~13, no.~8, pp. 1733--1745, 2014.

\bibitem{Ma-Zhang}
X.~Ma, J.~Zhang, X.~Yin, and K.~S. Trivedi, ``Design and analysis of a robust broadcast scheme for {VANET} safety-related services,'' \emph{IEEE Transactions on Vehicular Technology}, vol.~61, no.~1, pp. 46--61, 2012.

\bibitem{Zeng-Yu}
X.~Zeng, M.~Yu, and D.~Wang, ``A new probabilistic multi-hop broadcast protocol for vehicular networks,'' \emph{IEEE Transactions on Vehicular Technology}, vol.~67, no.~12, pp. 12\,165--12\,176, 2018.

\bibitem{Khan-Koubaa}
Z.~Khan and A.~Koubaa, ``Smartflow: An adaptive congestion avoidance protocol for smart transportation systems,'' in \emph{2020 International Wireless Communications and Mobile Computing (IWCMC)}, 2020, pp. 1535--1540.

\end{thebibliography}

\vfill

\end{document}